\documentstyle[12pt,openbib,epsf]{article}

\input epsf
\begin{document}
\vspace{1cm}
\begin{center}
{\large \bf Compatibility of a model for the QCD-Pomeron 
and chiral-symmetry breaking phenomenologies}\\[.4cm]
A.~A.~Natale~\footnotemark
\footnotetext{e-mail: natale@axp.ift.unesp.br} 
and P.~S.~Rodrigues da Silva~\footnotemark\\
\footnotetext{e-mail: fedel@axp.ift.unesp.br}
Instituto de F\'{\i}sica Te\'orica, Universidade Estadual 
Paulista\\
Rua Pamplona, 145, 01405-900, S\~ao Paulo, SP - Brazil
\end{center}
\thispagestyle{empty}
\vspace{1cm}

\begin{abstract}
The phenomenology of a QCD-Pomeron model based on
the exchange of a pair of non-perturbative gluons, {\it i.e.} gluon
fields with a finite correlation length in the vacuum, is studied
in comparison with the phenomenology of QCD chiral symmetry breaking,
based on non-perturbative solutions of Schwinger-Dyson equations for
the quark propagator including these non-perturbative gluon effects.
We show that these models are incompatible, and point out
some possibles origins of this problem. 
\end{abstract}
%
\newpage                                                                  
\section{Introduction}

The interpretation of the Pomeron in the framework of QCD is not fully      
understood. It is expected to be generated at least by two gluons            
exchange~\cite{low}. However, the exchange of two perturbative gluons cannot
reproduce the experimental results on diffractive scattering. In order to
circumvent such difficulty, Landshoff and Nachtmann~\cite{lan2}(LN) proposed a
model where the Pomeron is described by the exchange of two non-perturbative
gluons, whose properties are dictated by the expected structure of the QCD
vacuum. These non-perturbative gluons should not propagate over long distances, 
i.e. there is a finite correlation length for the gluon field in the vacuum,
which should be determined from first principles, and can be understood in terms
of gluon condensates~\cite{lan2}.                                          
The LN model describes successfully the experimental results of diffractive
phenomena using a quite simple ansatz for the non-perturbative gluon
propagator~\cite{don,cud}. Recently we improved the model using a QCD motivated
non-perturbative gluon propagator, where the correlation length is provided by a
dynamically generated gluon mass~\cite{hal}.                           
                                                                             
The remarkable success of this Pomeron model might provide physical insight in
the properties of the gluon propagator in the infrared region. It is, therefore,
natural to speculate on the ability of such a model to describe other low-energy
QCD phenomena, in particular, dynamical chiral symmetry breaking (DCSB). Within
the spirit of the LN model, we study the compatibility of both the Pomeron and
DCSB phenomenologies. The quark self-energy is computed through the numerical
calculation of the Schwinger-Dyson equation (SDE) for the quark propagator,
employing the same non-perturbative gluon propagators as the ones used
in the Pomeron phenomenology. According to
the traditional approach to the DCSB \cite{robwil}, 
these self-energies are used to obtain the experimental
values of the quantities that characterize the chiral breaking, such as the
quark condensate and the pion decay constant.

It is worth mentioning that the existence of a finite correlation length
for the gluon field was already proposed ten years before the LN 
model in a completely different context~\cite{cor}, and it gained
strong support by recent lattice simulation of the gluon 
propagator~\cite{bernard,marenzoni}. Therefore, a study of its
phenomenological implications is quite compelling, although
there are few places where this phenomenon can be effectively
tested.

Using only QCD motivated gluon propagators, {\it i.e.} propagators
determined from the solutions of Schwinger-Dyson equations or
obtained from lattice simulation, which satisfy the LN conditions
and are constrained by the diffractive scattering data, we show that they 
do not lead to satisfactory DCSB parameters.
In Section II we discuss some of the non-perturbative propagators
found in the literature, and show how they are constrained by 
the diffractive scattering data according to the LN model. In
Section III we use the non-perturbative gluon propagators, with the 
parameters found in the previous section, to compute the
relevant quantities for DCSB. In Section IV we discuss the incompatibility
found in the previous section and point out the possibles origins of
such a failure. 

\section{Diffractive scattering and gluon propagators}

There are several experimental features of diffractive scattering that
can be computed through the two-gluon exchange model of the Pomeron
according to the LN prescription, and here we will choose two simple
quantities which can be related to hadron-hadron total cross sections,
and show how they constrain the gluon propagators.
The first one will be the quark-Pomeron coupling and the second one, 
the pion-proton total cross section.

The strength of the Pomeron coupling to quarks $(\beta_0)$ at leading
order is given by~\cite{lan2}
\begin{equation}                                                            
\beta_0^2=\frac{1}{36\pi^2}\int_{}^{}\,d^2K                                 
\left[g^2(K^2)D(K^2)\right]^2,                                                   
\label{e1}                                                                  
\end{equation}                                                              
where $g^2(K^2)/4\pi$ is the running quark-gluon coupling, and the value of
$\beta_0^2=4$ GeV$^{-2}$ is extracted from the proton-proton
cross section. Notice that we will be working with Euclidean momenta
$(K^2=-k^2)$.

The amplitude of meson-nucleon scattering is~\cite{levin}
\begin{eqnarray}
A &=& \imath \frac{32}{9} s \int_{}^{} \, d^2K \,
[\alpha D(K^2)] [\alpha D((2{\bf Q}-{\bf K})^2)] \nonumber \\
&& \times 2 [f_M(Q^2)-f_M(({\bf Q} -
{\bf K})^2)] \nonumber \\
&& \times 3 \left[ f_N(Q^2) - f_N \left( Q^2 - \frac{3}{2}
{\bf Q.K} + \frac{3}{4} K^2 \right) \right],
\label{e2}
\end{eqnarray}
where $s$ is the square of the center-of-mass energy, to the
couplings $(\alpha)$ we associate the same momentum of its
multiplying propagator, $f_M$
and $f_N$ are respectively the meson and nucleon form factors.
The total cross section is related to this amplitude by
\begin{equation}
\sigma_T = \frac{Im A(s,t=0)}{s},
\label{e3}
\end{equation}
and Eq.(\ref{e2}) is computed with the form factors in the 
pole approximation
\begin{equation}
f_i (K^2) = \frac{1}{\left( 1 + \frac{<r_i^2>}{6} K^2 \right) }.
\label{e4}
\end{equation}

We will calculate the pion-proton total cross section. Actually, 
the two gluon exchange model cannot describe the full cross
section growth with energy $( \propto s^{0.0808})$, which
appears in a recent fit for the $\pi - p$ total cross 
section~\cite{dola}
\begin{equation}
\sigma_T^{\pi p} = 13.63 s^{0.0808} + 36.02 s^{-0.4525}.
\label{e5}
\end{equation}
The model can only accommodate the value $13.63$ in the above
expression. We used the following mean squared 
radii for the proton and pion, respectively~\cite{povh}: $<r_p^2>=0.67$ fm$^2$
and $<r_{\pi}^2>=0.44$
fm$^2$.
Inspection of Eq.(\ref{e1}) and Eq.(\ref{e2}) show that they
are quite dependent on the gluon propagator expression at
$k^2 \rightarrow 0$ and, therefore, 
well suited to study its infrared behavior.

We will now consider the propagators that satisfy some of the
basic assumptions of the LN model: (a) the propagator has
a finite correlation length and (b) it is finite at $k^2=0$.
We will only deal with QCD motivated propagators,
{\it i. e.} those obtained as solutions of the 
SDE for the gluon polarization
tensor, as well as obtained in a lattice simulation.
We start with the propagator determined by Cornwall~\cite{cor}, 
which has already been used by one of us to describe
some diffractive scattering processes~\cite{hal},
and has the following expression  
\begin{equation}
D_c^{-1}(K^2) = \left[K^2+m^2(K^2)\right]
bg^2 \ln \left[ \frac{K^2+4m^2(K^2)}
{\Lambda^2} \right],                                                        
\label{e6}                                                                  
\end{equation}                                                              
where $m^2(K^2)$ is a momentum-dependent dynamical mass 		    
\begin{equation}                                                   
m^2(K^2) = m_g^2 \left[
\frac{\ln\left(\frac{K^2+4m_g^2}
{\Lambda^2}\right)}{\ln\frac{4m_g^2}        
{\Lambda^2}}
\right]^{-12/11}.                                                           
\label{e7}                                                                  
\end{equation}								    
In the above equation $m_g$ is the effective gluon mass,                    
and $b=(11N-2n_f)/48\pi^2$ is the leading order coefficient                  
of the $\beta$ function of the renormalization group equation, 
with $N=3$ for QCD, and where 
$n_f$ is the number of flavors
taken to be 3. In Eq.(\ref{e6}), the quark-gluon coupling strength $g$ is 
such that $g\simeq 1.5-2$, which was determined from a fit of Eq(\ref{e6}) 
to the numerical solution of the SDE for the gluon propagator.
For the running coupling constant $g(K^2)$ we assume a functional form
which interpolates between a constant and the renormalization group
asymptotic behavior 
\begin{equation}
\alpha_s (K^2) \equiv \frac{g^2(K^2)}{4\pi} = \frac{12\pi / 
(33 -2n_f)}{\ln(\frac{K^2}{\Lambda^2}+\tau)},
\label{e8}
\end{equation}
where $\tau= \kappa^2/\Lambda^2$, and we chose $\kappa = m_g$. Such freeze-out 
of the coupling constant in the infrared region is consistent with the
Cornwall and Papavassiliou~\cite{cjp} study of the trilinear gluon vertex.
A similar form for the coupling constant has also
been used in many phenomenological applications (see,
for instance, Ref.~\cite{roberts} and references therein).

It has also been pointed out that the dynamical gluon mass
may have a faster decrease with the momentum~\cite{corhou},
according to the operator product expansion (OPE) determination
of the ultraviolet behavior of the gluon polarization tensor~\cite{lavelle}
\begin{equation}
\Pi_{UV} (k^2) \sim \frac{34N \pi}{9(N^2-1)}
\frac{\left\langle \alpha_s G^2 \right\rangle}{k^2},
\label{e9}
\end{equation}
where $\left\langle \alpha_s G^2 \right\rangle$ is the gluon
condensate. Therefore, to be consistent with the massive
Cornwall propagator and with OPE, we will consider a gluon
self-energy that interpolates between the constant
infrared behavior of Eq.(\ref{e7}) and the ultraviolet
one of Eq.(\ref{e9}) which will be given by
\begin{equation}
m^2 (K^2) = \mu^2 \theta(\mu^2-K^2) + \frac{\mu^4}
{K^2} \theta(K^2-\mu^2),
\label{e10}
\end{equation}
where the scale $\mu^2$ will be limited by Eq.(\ref{e1})
and Eq.(\ref{e2}).

Another infrared finite propagator has been found by
Stingl and collaborators as a solution of the SDE for
the gluon polarization tensor~\cite{sting}. Its form agrees
with that derived by Zwanziger based on considerations
related to the Gribov horizon~\cite{gribov}, and is given
by
\begin{equation}
D_s(K^2)= \frac{1}{K^2 + \frac{\mu_s^4}{K^2}},
\label{e11}
\end{equation}
where $\mu_s$ is a mass scale not determined in 
Ref.~\cite{sting}. The term $\mu_s^4/K^2$ is exactly
what is expected by OPE analysis whenever a mass scale
for the gluon is introduced, and the full solution found
in Ref.~\cite{sting} did contain a mass term, although
its consequences were not pursued. It is also interesting
to notice that the Bernard {\it et al.} lattice result
for the gluon propagator~\cite{bernard} can be fitted by 
Eq.(\ref{e6}) as well as Eq.(\ref{e11}).

Finally, Marenzoni {\it et al.}~\cite{marenzoni} 
also performed a lattice
study of the gluon propagator in the Landau gauge, 
obtaining for its infrared behavior the following fit
\begin{equation}
D_m(K^2)= \frac{1}{m_l^2 + Z K^2 (\frac{K^2}{\Lambda^2})^\eta},
\label{e12}
\end{equation}
where $m_l$, $Z$ and $\eta$ are constants determined by 
the numerical simulation. $m_l$ is of 
${\cal O}(\Lambda \approx 160 \, MeV)$, 
$Z \approx 0.4$ and $\eta \approx 0.5$ 
what is slightly different from the previous propagators.
The results of Bernard {\it et al.} also show the behavior
$(K^2)^\eta$, but with a smaller value for $\eta$. In this
case all the propagator parameters are determined, and we
simply have to see if they are consistent with diffractive
scattering and DCSB. All the above propagators where 
obtained in Landau gauge, except Cornwall's one, whose
massive solution was shown to be gauge invariant.

Performing the calculation of the integrals in Eqs. (\ref{e1}) 
and (\ref{e2}) for each one of the above propagators, we 
obtained the curves for $\beta_0^2$ displayed in Fig.(1), 
and the pion-proton total cross section shown in Fig.(2) as 
functions of the gluon mass.
From these figures and the experimental values of $\beta_0$, 
and the pion-proton total cross section, it was possible 
to establish an effective gluon mass for each propagator. Our 
results are shown in Table \ref{table1}.

It is clear from this table that almost all the gluon
mass scales that fit the experimental value of 
$\beta_0$ give reasonable (within $10\%$) values for 
the total pion-proton cross section, 
except in the case of the lattice propagator, Eq.(\ref{e12}), where the
comparison with the experimental values 
yields effective gluon masses differing by $35\%$. It is evident
that not only the mass scale plays a role in this phenomenology,
but also the functional form of the propagator introduces
differences in the calculation, and this is why this problem
becomes more interesting, because we can speculate about the
infrared behavior of the gluon propagator.
As we shall see in the next section, we will fail to obtain
DCSB with the above propagators for the gluon mass scales
presented in Table \ref{table1}.

\section{DCSB with LN type propagators}

We discuss dynamical chiral symmetry breaking                         
following the traditional approach \cite{robwil}.                           
This consists in solving the Schwinger-Dyson equation
(SDE) for the quark propagator and look for a 
mass term dynamically generated 
in this propagator. 

The Schwinger-Dyson equation for the quark propagator is
\begin{equation}
S^{-1}(p)=\not\!{p}
-\imath \frac{4}{3} \int_{}^{} \frac{d^4q}{(2 \pi )^4}
\gamma_\mu S(q)\Gamma_\nu(p,q)g^2D^{\mu\nu}(p-q) ,
\label{e13}
\end{equation}
where we write the gluon propagator in the form
\begin{equation}
g^2D^{\mu \nu}(q)= \frac{4\pi \alpha (-q^2/\Lambda^2)}{q^2}
\left( -g^{\mu \nu}+\frac{q^{\mu}q^{\nu}}{q^2} \right).
\label{e14}
\end{equation}
The propagator has been written  
in the Landau gauge, which will be used throughout our work.
In the above equations $\Gamma_\nu (p,q)$ is the vertex function, and
$\alpha(-q^2/\Lambda^2)$ is the QCD running coupling constant given
by Eq.(\ref{e8}). 

To proceed further we also need to introduce an ansatz for
the quark-gluon vertex $\Gamma^\mu (p,q)$, which must satisfy a
Slavnov-Taylor identity that, when we neglect ghosts, reads
\begin{equation}
(p-q)_\mu \Gamma^\mu (p,q)=S^{-1}(p)-S^{-1}(q).
\label{e15}
\end{equation}
This identity constrains the longitudinal part of the 
vertex, and if we write $S^{-1}(p)$ in terms of scalar
functions
\begin{equation}
S^{-1}(p)=A(p) \not\!{p} - B(p), 
\label{e16}
\end{equation}
we find the solution~\cite{krein}
\begin{eqnarray}
\Gamma^\mu (p,q) &=& \frac{(p-q)^\mu}{(p-q)^2}
\left( [A(p^2)-A(q^2)] \not\!{q}  
- [B(p^2)-B(q^2)] \right) \nonumber \\
&& + A(p^2)\gamma^\mu +  \,\, transverse \, part,
\label{e17}
\end{eqnarray}
which is a much better approximation than the use of the bare 
vertex \cite{CUPE}. Assuming that the transverse vertex part vanishes 
in the Landau gauge we obtain
\begin{equation}
D^{\mu \nu}(p-q)\Gamma_\nu (q,p)=D^{\mu \nu}
(p-q)A(q^2)\gamma_\nu,
\label{e18}
\end{equation}
and arrive at the approximate Schwinger-Dyson equation
\begin{equation}
[A(p^2)-1]\not\!{p}-B(p^2) = \imath\frac{4}{3} \int 
\frac{d^4q}{(2 \pi )^4} 
g^2 D^{\mu \nu}(p-q)\gamma_\mu 
\frac{A(q^2)}{A(q^2)\not\!{q}-B(q^2)}
\gamma_\nu.
\label{e19}
\end{equation}
Going to Euclidean space, we will be working with the following
nonlinear coupled integral equations for the quark wave-function
renormalization and self-energy
\begin{eqnarray}
[A(P^2)-1]P^2 &=& \frac{16\pi}{3} \int_{}^{} \frac{d^4Q}{(2\pi)^4}
\frac{\alpha ((P-Q)^2/\Lambda^2)}{\Phi[(P-Q)^2]} \nonumber \\
&& \times \left( P.Q +2 
\frac{P.(P-Q)Q.(P-Q)}{(P-Q)^2} \right) \nonumber \\
&& \times \frac{A^2(Q^2)}{A^2(Q^2) Q^2+B^2(Q^2)},
\label{e20}
\end{eqnarray}
\begin{equation}
B(P^2)=16\pi \int_{}^{} \frac{d^4Q}{(2\pi)^4}
\frac{\alpha ((P-Q)^2/\Lambda^2)}{\Phi[(P-Q)^2]}
\frac{A(Q^2)B(Q^2)}{A^2(Q^2)Q^2+B^2(Q^2)},
\label{e21}
\end{equation}
where $Q^2=-q^2$ and $P^2=-p^2$, and we introduced a function
$\Phi[(P-Q)^2]$ which, in the case of the perturbative
propagator, is simply $\Phi[(P-Q)^2]=(P-Q)^2$, for a
massive bare gluon it will have the form 
$\Phi[(P-Q)^2]=(P-Q)^2+m_g^2$, and will be more
complex expression according to the propagators we
discussed in the previous section.

The numerical code we used to solve the above equations is the
same of Ref.~\cite{WKR}, and for each one of the propagators
in Section II we substitute: (a) $\Phi (K^2)=D_c^{-1}(K^2)$,
(b) $\Phi (K^2)=D_{cm}^{-1}(K^2)$,
(c) $\Phi(K^2)=D_s^{-1}(K^2)$ and (d) $\Phi(K^2)=D_m^{-1}(K^2)$.
After obtaining for each case the function $A(p^2)$ and $B(p^2)$,
we can compute the quark condensate, which is expressed as
\begin{equation}
\left\langle \bar{q} q \right\rangle = - 12 \imath \int_{}^{\Lambda}
\frac{d^4p}{(2\pi)^4} \, \frac{Z(p^2) M(p^2)}{p^2-M^2(p^2)},
\label{e22}
\end{equation}
where $Z^{-1}(p^2)=A(p^2)$ and $M(p^2)=B(p^2)/A(p^2)$, and we
will also calculate the pion decay constant, which in terms of
the functions $A$ and $B$ of Eq.(\ref{e20}) and Eq.(\ref{e21})
is given by
\begin{eqnarray}
f_{\pi}^2 &=& - 12 \imath \int \frac{d^4p}{(2\pi)^4}
\frac{AB}{(A^2p^2 - B^2)^2}  [ AB 
\left( 1+\frac{p^2}{2} \frac{d \ln A}{dp^2} \right) \nonumber \\
&& + \frac{p^2}{2} 
\left( B \frac{dA}{dp^2} - A \frac{dB}{dp^2} \right) 
\left( 1 + p^2 \frac{d \ln A}{dp^2} \right) ] .
\label{e23}
\end{eqnarray}

For all the propagators discussed here, with the
respective gluon mass scales given in Table \ref{table1} ,
we have not obtained dynamical quark mass 
generation! Our results for $f_{\pi}$ and $\langle\bar{q} q\rangle$
are identically zero (at least ten orders of magnitude below
the scale $\Lambda$). The gluon mass scales of Table \ref{table1}
are too large, in the sense that they cause a too strong screening
of the force necessary to generate the symmetry breaking. To illustrate
this incompatibility we present in Fig.(3) the curve of the 
dynamically generated quark mass $(M(p^2=0))$ against the gluon mass,  
in the case of the Cornwall propagator (Eq.(\ref{e6})), which gives the
largest signal of mass generation that we obtained among all the
propagators described above. Notice that in Fig.(3) to obtain
dynamical quark masses of ${\cal O}(300~\, MeV)$, we would need 
a gluon mass almost half the value necessary to satisfy the
constraints from diffractive scattering. 

The inconsistency between diffractive 
scattering and the chiral symmetry breaking
phenomenologies is not only a matter of adjustment of the infrared
gluon mass scale, this scale really plays a different role in
both cases. We need large gluon masses $(m_g >> \Lambda)$ to soften the t-channel
singularity in diffractive scattering, but smaller gluon masses
if we do not want to erase the dynamical quark mass generation.
It is obvious that we have an inconsistency, although
it is far from clear which is the solution. Nevertheless, this comparison of phenomenologies gives
a powerful tool to constrain the detailed behavior of the infrared
gluon propagator. 

\section{Conclusions}

The LN Pomeron model is able to explain a large amount of
experimental data on diffractive scattering~\cite{cud,hal},
making use of an infrared finite gluon propagator. As we have
seen in Section 2, we need only a unique gluon mass scale to
fit the pomeron-quark coupling and the pion-proton total cross
section. On the other hand we have an extensive and successful
phenomenology of DCSB, which was performed
mostly with a ``perturbative'' $(1/k^2)$ gluon propagator.
In Section 3 we followed the standard procedures to
compute the dynamical quark mass, quark condensate and
pion decay constant, with the same non-perturbative
propagators prescribed by the LN model. Both phenomenologies
are inconsistent. We need large gluon masses for diffractive
scattering, and small ones to generate appropriate chiral symmetry
breaking parameters!

There is a clear difference between our calculation of
DCSB with the previous ones. The use of ``massive'' propagators
clearly screens the force necessary to generate the chiral
symmetry breaking, and it is probably in this direction
that we may look for troubles, another possibility would
imply that the LN model is incorrect, or that the
standard phenomelogy of DCSB has nothing to do with a
finite correlation length for gluons. Let us discuss
about each one of these possibilities. The first
common point to both phenomenologies is the existence of a
mass scale for the gluon propagator, however, 
there is another claimed form for the gluon propagator in the infrared
region, namely, $1/k^4$ (see \cite{pen} and references therein).
In view of the theoretical arguments of 
Ref.~\cite{cor,lavelle,sting,gribov} about gluon mass scales,
and the numerical simulations on the
lattice~\cite{bernard,marenzoni}, we were compelled
to assume that the gluon field definitively has a finite correlation
length in the vacuum, discarding the $1/k^4$ solution. 

The gluon propagator being infrared finite,
there is no comparison with the experimental data which would
force us to abandon the LN model, and it remains to see if the
DCSB phenomenology with such propagator is still consistent.
Another difference to standard DCSB calculations is the use of
the QCD running coupling constant given by Eq.(\ref{e8}), which,
as we said previously, is consistent with a study of the 
trilinear gluon vertex, and naturally embodies the freezing
of the coupling at the relevant gluon mass scale~\cite{cjp}.
Both phenomenologies, diffractive
scattering and DCSB, depend strongly on the behavior of this
coupling in the infrared, however, its effect works basically
in the same direction, i.e. increasing the coupling in the
infrared we increase the diffractive scattering parameters as well
as the dynamical quark mass. Therefore, there will not be much room to
changes in this direction. Finally, if we assume that the
LN Pomeron model is correct, the gluon field has a finite
correlation length in the vacuum, and the coupling constant
has the freezing-out referred to above, we are forced to
say that the dynamical quark mass is not generated by
single-gluon forces, and the usual calculation of dynamical quark
masses should be modified when the gluon mass is taken into
account. Actually, this view is not new, studying the coupled
fermion gap and vertex equations for DCSB in QCD it was
also found an incompatibility~\cite{JPC}, and it was argued that the
standard technique of using one-gluon exchange in the SDE is not
suitable when we consider the effect of massive gluons, and our
work, in a different way, corroborates their result.
 
We finalize stressing that we are still far from knowing all
the subtleties of QCD, and we have formalized the incompatibility
between a QCD Pomeron model and DCSB. 
We believe that the confront of these different phenomenologies
may provide strong constraints on the infrared behavior of the
gluon propagator. 

\section*{Acknowledgments}
This research was partially supported by the Conselho Nacional de
Desenvolvimento Cientifico e Tecnologico (CNPq)(AAN), and Funda\c c\~ao
de Amparo a Pesquisa do Estado de S\~ao Paulo (FAPESP)(PSRS).  We would like             to thank Jean-Rene Cudell and Francis  Halzen for
their comments on a previous manuscript, and also G. Krein for discussions,
and for helping us with the numerical code of Ref.\cite{WKR}. We are 
grateful for the kind hospitality at the Institute for Elementary 
Particle Physics Research, University of Wisconsin - Madison (AAN), and at
the International Centre for Theoretical Physics (ICTP) (PSRS), where
part of this work was done.

\newpage
\section*{Figure Caption}

\noindent
{\bf Table~1} Gluon masses (in MeV) obtained fitting 
$\beta_0$ and $\sigma_{\pi p}$ for the several propagators 
discussed above. We denote by $D_{cm}$ the gluon propagator
given by Eq.(\ref{e6}) with the dynamical gluon mass 
given by Eq.(\ref{e10}).
$\\$
$\\$
\noindent
{\bf Fig.~1} Curves for $\beta_0^2$ as a function of the gluon mass. We 
constrain the gluon mass by fixing $ \beta_0^2=4.0 $GeV$^{-2}$. Each 
gluon propagator is represented by:  dashed line ($D_c$), dash-dotted 
line ($D_m$), dotted line ($D_{s}$), solid line ($D_{mc}$).   
$\\$
$\\$
\noindent
{\bf Fig.~2} Curves for pion-proton total cross section as a function of the 
gluon mass. The constraint on the gluon mass comes from fixing the value
 $\sigma_{\pi p} = 13.63$ mb. The gluon propagators are represented by:  
dashed line ($D_c$), dash-dotted line ($D_m$), dotted line ($D_{mc}$), solid 
line  ($D_s$).
$\\$
$\\$
\noindent
{\bf Fig.~3} Dynamically generated quark mass for the Cornwall's gluon propagator as a function of the gluon mass.
\begin{table}
\begin{center}
\begin{tabular}{|l|c|r|}\hline\hline
\textbf{Propagator} & \textbf{$\beta_0$} & \textbf{$\sigma_{\pi p}$} \\ \hline\hline
$D_c$ & 760 & 779 \\ \hline
$D_{cm}$ & 513 & 472 \\ \hline
$D_s$ & 422 & 387 \\ \hline
$D_m$ & $ 525 $ & 388 \\ \hline
\end{tabular}
\end{center}
\caption{}
\label{table1}
\end{table}
\begin{figure}[htb]
\epsfxsize=0.6\textwidth
\begin{center}
\leavevmode
\epsfbox{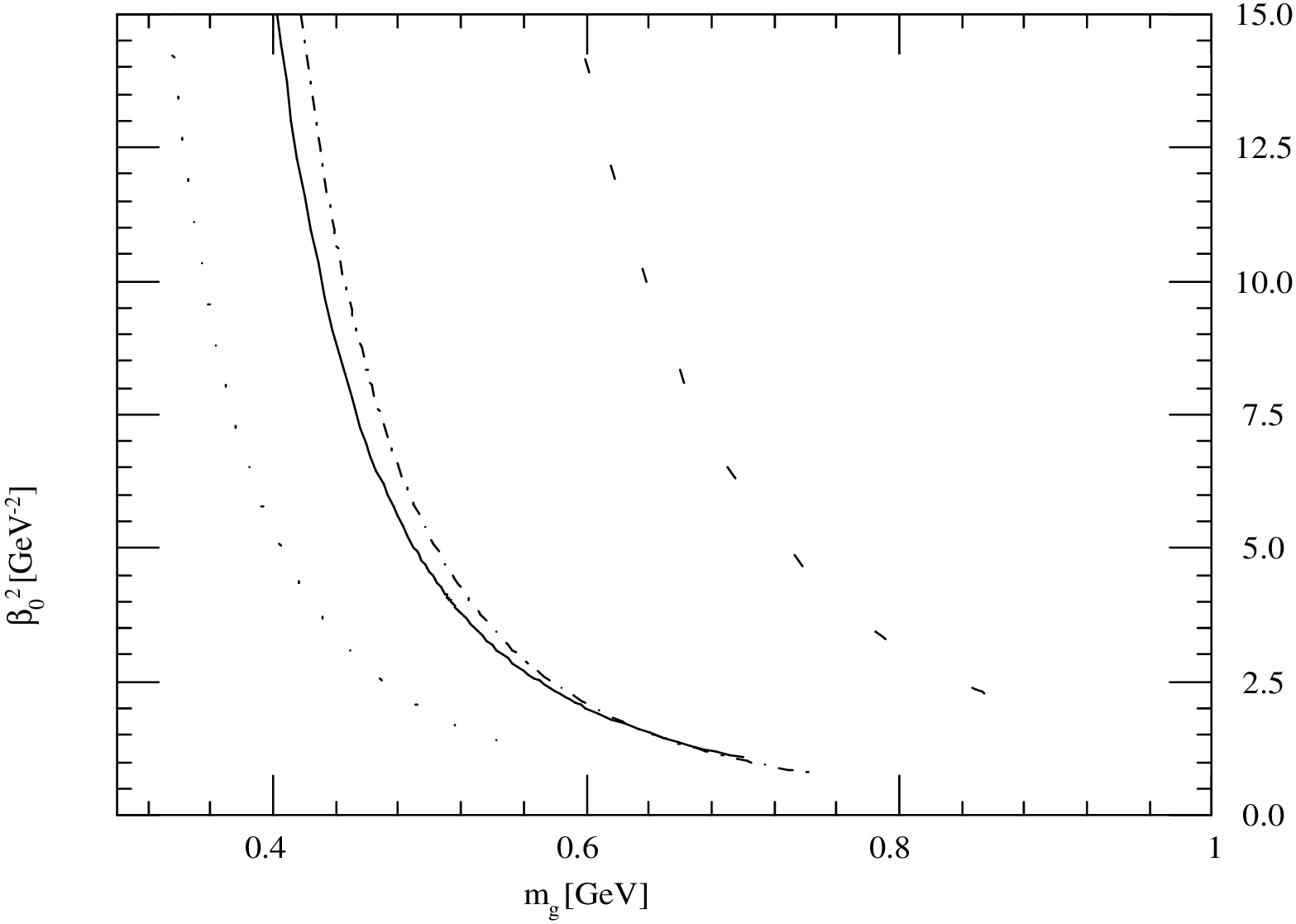}
\end{center}
\caption{}
\label{f1}
\end{figure}
\begin{figure}[htb]
\epsfxsize=0.6\textwidth
\begin{center}
\leavevmode
\epsfbox{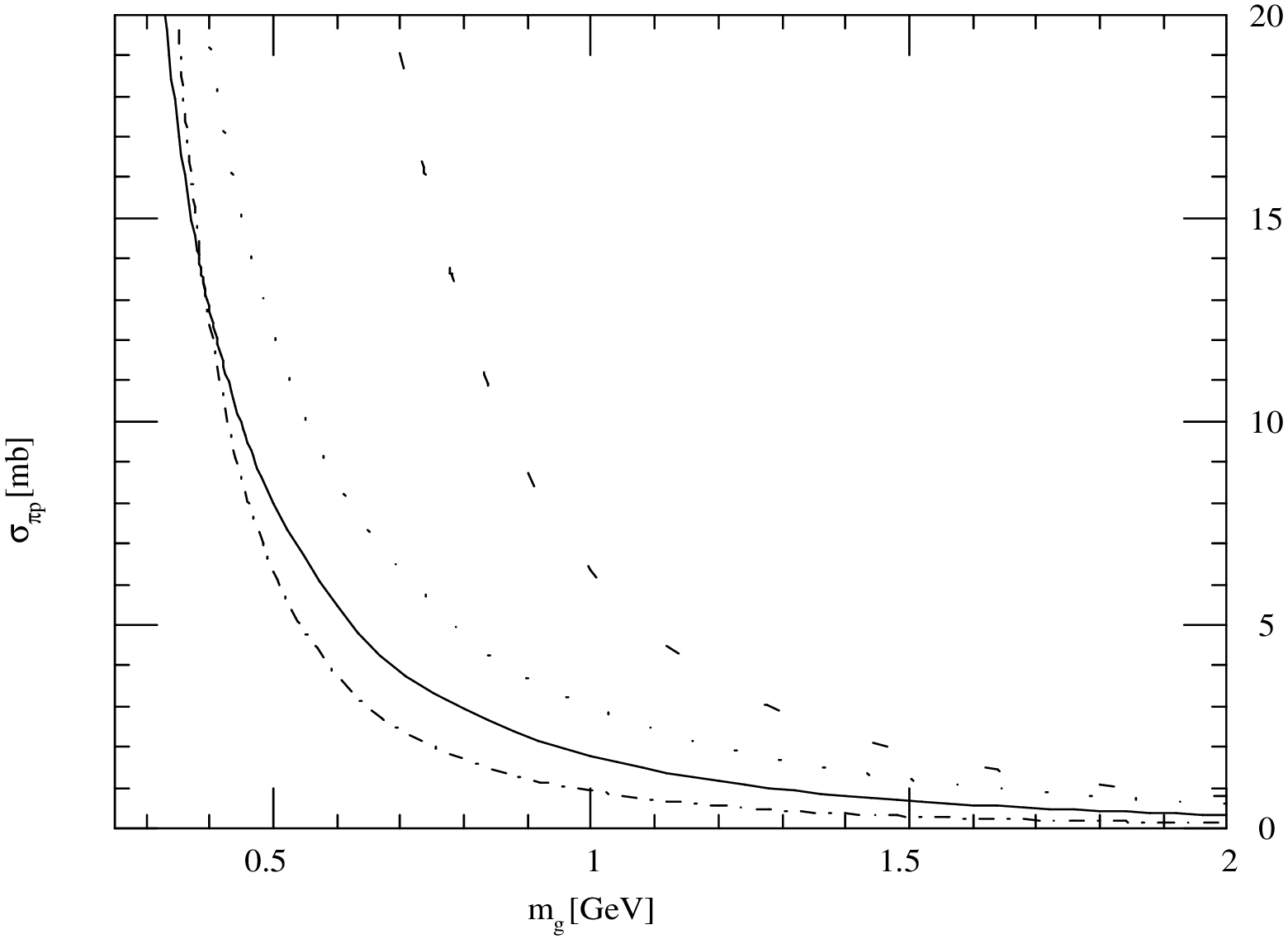}
\end{center}
\caption{}
\label{f2}
\end{figure}
\begin{figure}[htb]
\epsfxsize=0.6\textwidth
\begin{center}
\leavevmode
\epsfbox{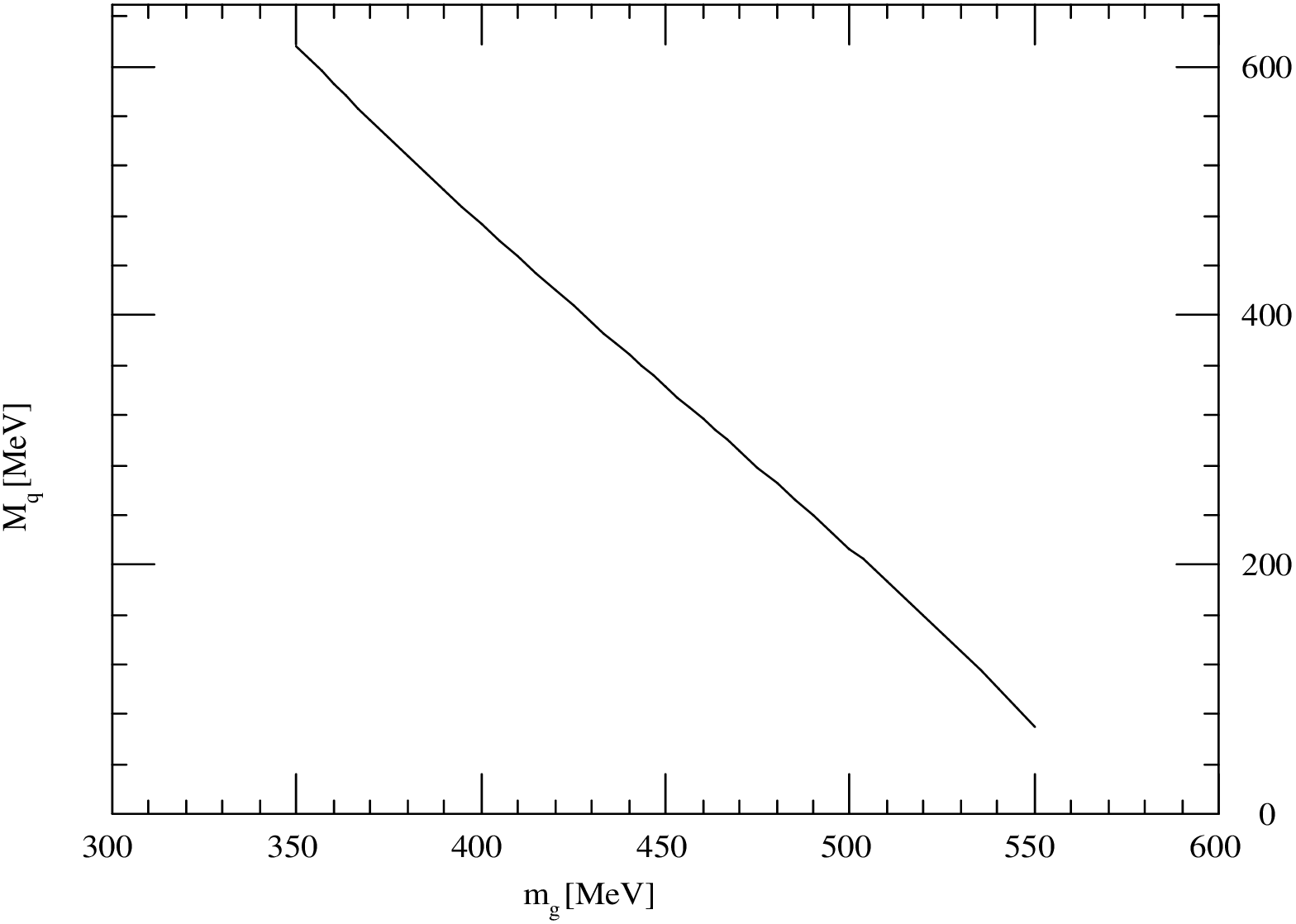}
\end{center}
\caption{}
\label{f3}
\end{figure}

\end{document}